\documentclass[12pt]{iopart}
\begin{document}
\hyphenation{Schwarz-schild}
\title[2+1 D Lattice Universes]{Lattice Universes in 2+1-dimensional gravity}

\author{Dieter R. Brill}  

\address{Department of Physics, University of Maryland, College Park, MD 20742, USA\\email: brill@physics.umd.edu}

\begin{abstract}
Lattice universes are spatially closed space-times of spherical topology in the large, containing masses or black holes arranged in the symmetry of a regular polygon or polytope. Exact solutions for such spacetimes are found in 2+1 dimensions for Einstein gravity with a non-positive cosmological constant. By means of a mapping that preserves the essential nature of geodesics we establish analogies between the flat and the negative curvature cases. This map also allows treatment of point particles and black holes on a similar footing.
\end{abstract}




\section{Introduction}
On a spacetime safari, Vince has taught us, it is not enough to go from one big game to another, it also pays to attend to the smaller details. Here I give some of the details in 2+1-dimensional gravity.

Lattice universes in (3+1) dimensions were introduced by Lindquist and 
Wheeler \cite{LW} as 
empty space approximations to Friedmann universes, in order to show that empty curved spacetime can act like a matter source: instead of 
uniformly distributed matter, these models contain black holes, and are 
therefore purely ``geometrodynamical" in the sense that the sourceless Einstein 
equations hold everywhere. To approximate a uniform, spherical universe, space is divided into a lattice of equal cells,  
each cell is dominated by a Schwarzschild geometry, and the cells are those of a 3-dimensional regular polytope.  Because the continuous
spherical symmetry of the Schwarzschild geometry disagrees with the discrete
symmetry of the cells, these fit together only approximately. Lindquist and Wheeler give various criteria by which the quality of the approximation 
may be judged, but regret that there is no exact solution with a finite 
number of particles, with which the approximate solution may be compared. A comparison with a dust-filled Friedmann universe shows that, in appropriate respects, the regular polytope 
approximation to a sphere improves with the number of ``particles" (three-dimensional faces of the polytope, which themselves are the interiors of regular polyhedra), where the largest number of ``particles" possible in this approximation is 600.

A similar construction may be performed in 2+1 dimensional sourceless Einstein 
theory with a negative cosmological constant, since such spaces admit black hole  (BTZ) geometries \cite{BTZ}. In this case all exact solutions are locally isometric to anti-deSitter (AdS) space and differ only by their global topology, so at least in 
principle they are known. Indeed one knows how to construct 
2+1-dimensional multi-black-hole geometries and to characterize them uniquely by certain parameters, which describe the arrangement of the black holes and their masses \cite{B3}. But it is far from obvious what values of such 
parameters correspond, for example, to the most nearly uniform 
distribution of the black holes on a ``background" 2-sphere. In this paper 
we will outline a construction procedure appropriate for such cosmologies, with parameters of clear physical meaning. In distinction to the Lindquist-Wheeler paper (written before the discovery of the Kerr metric) we will also consider universes containing a uniform distribution of black holes with angular momentum.

\section{Time-symmetric Universes Modeled on Regular Polyhedra}
The simplest kind of lattice universe is a time-symmetric one. (This is the only type discussed by Lindquist and Wheeler; for a generalization see Brill et al \cite{BCI}.)
Time-symmetry implies that there is an ``initial" 2-dimensional spacelike Cauchy surface of 
vanishing extrinsic curvature. The intrinsic curvature of the surface must then
be constant negative, like that of the surrounding locally AdS spacetime. 
The Cauchy data for a single black hole are known from the BTZ solution, which has 
the usual wormhole shape that is symmetrical about a throat. We will use the presence of such a throat, or of
an apparent horizon (which we sometimes call simply a horizon) to 
identify a region containing a black hole. This seems the best one can do in a collapsing universe. These black holes, as seen from 
the interior of the universe, are not 
surrounded by real horizons in the usual sense, related to null infinity, because all null geodesics end at the collapse singularity. (Typically these 2+1-dimensional spacetimes and their null geodesics can be continued across the singularity, but because the continuation is considered unphysical, for example because of the presence of closed timelike curves, the manifold is taken to end at the singularity. Null geodesics starting inside a horizon and directed ``into" the black hole may continue to a null infinity, if there is such on the ``other" side of such a throat, where there will then be a genuine horizon; but this horizon is not relevant to causal physics of the finite interior universe, and in any case multi-black-holes always appear only as structures {\it inside} that horizon.)

\begin{figure}
\unitlength 1.00mm
\linethickness{0.4pt}
\begin{picture}(63.00,70.00)(-30,0)
\bezier{72}(29.65,52.00)(36.00,58.00)(42.35,52.00)
\bezier{72}(42.35,52.00)(47.79,45.00)(42.35,38.00)
\bezier{72}(29.65,52.00)(24.21,45.00)(29.65,38.00)
\bezier{72}(29.65,38.00)(36.00,32.00)(42.35,38.00)
\bezier{72}(48.76,58.50)(53.70,56.40)(48.76,55.04)
\bezier{72}(37.24,58.62)(32.30,56.83)(37.24,55.04)
\bezier{72}(37.24,55.04)(43.00,53.51)(48.76,55.04)
\bezier{72}(46.91,52.35)(51.00,58.65)(55.09,52.35)
\bezier{72}(46.91,52.35)(43.40,45.00)(46.91,37.65)
\bezier{72}(46.91,37.65)(51.00,31.35)(55.09,37.65)
\bezier{72}(34.00,32.17)(35.60,35.70)(43.00,35.98)
\bezier{72}(43.00,35.98)(50.40,35.70)(51.23,33.17)
\bezier{64}(41.17,70.00)(40.17,58.83)(35.83,58.17)
\bezier{56}(35.83,58.17)(30.00,55.50)(27.50,48.67)
\bezier{64}(44.17,70.00)(45.17,58.83)(49.50,58.17)
\bezier{64}(41.17,19.17)(40.17,29.33)(35.83,31.00)
\bezier{64}(44.17,19.17)(45.17,30.33)(49.50,32.00)
\bezier{60}(35.50,31.17)(28.33,35.50)(27.33,42.33)
\bezier{52}(49.67,32.17)(55.33,35.50)(56.50,41.50)
\bezier{52}(63.00,43.50)(54.33,41.33)(53.00,37.67)
\bezier{52}(63.00,46.50)(54.33,48.67)(53.00,52.33)
\bezier{44}(50.50,57.67)(54.33,55.83)(56.17,49.17)
\bezier{12}(27.90,46.56)(29.00,47.81)(30.23,46.56)
\bezier{12}(30.23,46.56)(31.21,45.16)(30.23,43.59)
\bezier{12}(30.23,43.59)(29.00,42.03)(27.78,43.44)
\bezier{12}(27.78,43.44)(26.80,45.00)(27.90,46.56)
\bezier{28}(29.33,47.33)(32.83,48.17)(34.50,51.67)
\bezier{28}(29.33,42.67)(32.83,41.83)(34.50,38.33)
\end{picture}
\vskip-20\unitlength
\begin{caption}{Qualitative character of the space in a lattice universe.}
\end{caption}
\end{figure}

The part near the throat of a single black hole geometry can be embedded
in Euclidean space as a pseudosphere, for example the figure of 
revolution of a 
tractrix.\footnote{The surface of revolution of that particular curve has the spacelike geometry of a 
$m=0$ BTZ black hole, whose throat is at infinity.} Following Lindquist and Wheeler we now place pseudospheres on a sphere, centered above each vertex of some regular inscribed Euclidean polyhedron, such as an octahedron, and in such a way that the pseudospheres are tangent to the sphere and touch each 
other at isolated points. This is shown in Figure 1, a figure drawn after 
Fig.~3 in the classic work of Lindquist and Wheeler.\footnote{Here the 
polyhedron that specifies the lattice structure, as used by Lindquist and Wheeler, is conjugate to the octahedron, namely a cube.} This surface allows us to visualize accurately the momentary, spacelike geometry of a 2+1-dimensional black hole universe. However, it only approximates sourceless initial values, because parts of the 
2-sphere remain uncovered by the pseudospheres. These regions have 
positive curvature rather than the negative curvature required for a 
time-symmetric surface in locally AdS space. (The positive curvature 
could be provided by matter in the gap regions. Because the pseudospheres 
are tangent to the sphere, the boundaries are joined sufficiently 
smoothly so that the curvature and the necessary matter density has only a 
jump discontinuity, and no surface matter density is required at the 
boundary. Alternatively the gap could be filled with further 
pseudospheres, of which an infinite decreasing sequence would be required to fill the gap completely.)

\subsection{Universes containing point particles}

In 2+1 dimensions we can construct an {\em exact} initial 
geometry, which has constant curvature everywhere. Consider first the case $\Lambda = 0$, where time-symmetric surfaces have vanishing curvature. Any regular polyhedron, such as the tetrahedron of Fig.~2a, consists of flat faces, joined at edges where the surface is also {\em intrinsically} flat.  At the vertices there is however a deficit angle compared to flat space (of size $4\pi/N$ for a regular solid with $N$ vertices). In 2+1 dimensions such a conical singularity signifies the location of a point particle, of mass equal to the angle deficit \cite{DJT}. So the tetrahedron is a model universe of 
approximately spherical shape, but an exact initial geometry of 
2+1-dimensional $\Lambda = 0$ Einstein theory, containing four equal 
point masses. Similarly, the surface of any static $N$-vertex polyhedron that remains unchanged in time can be considered a 2+1-dimensional $N$-particle universe, for the time-development of an initial polyhedron is simply that polyhedron continued statically in time. The total mass of any such closed universe, computed as the sum of the particle masses, is always $4\pi$ \cite{DJT}.

\begin{figure}
\unitlength 1.00mm
\linethickness{0.4pt}
\begin{picture}(141.00,45.00)
\put(30.00,7.00){\line(-5,2){16.00}}
\put(30.00,7.00){\line(1,5){2.00}}
\thicklines
\put(25.00,27.00){\line(1,-4){5.00}}
\put(14.00,13.40){\line(4,5){10.88}}
\put(32.00,17.00){\line(-2,3){6.90}}
\thinlines
\bezier{20}(14.00,13.50)(23.00,15.25)(32.00,17.00)
\put(25.00,27.00){\makebox(0,0)[cb]{A}}
\put(14.00,13.00){\makebox(0,0)[rc]{B}}
\put(30.00,7.00){\makebox(0,0)[ct]{C}}
\put(32.00,17.00){\makebox(0,0)[lc]{D}}
\put(45.00,7.00){\line(1,0){40.00}}
\put(85.00,7.00){\line(-3,5){20.00}}
\put(45.00,7.00){\line(3,5){20.00}}
\put(65.00,7.00){\line(-3,5){10.00}}
\put(65.00,7.00){\line(3,5){10.00}}
\put(55.00,23.67){\line(1,0){20.00}}
\put(65.00,7.00){\makebox(0,0)[ct]{C}}
\put(55.00,24.00){\makebox(0,0)[rb]{B}}
\put(75.00,24.00){\makebox(0,0)[lb]{D}}
\put(45.00,7.00){\makebox(0,0)[rc]{A}}
\put(85.00,7.00){\makebox(0,0)[lc]{A}}
\put(65.00,40.00){\makebox(0,0)[cb]{A}}
\thicklines
\put(100.00,12.00){\line(1,0){20.00}}
\put(120.00,12.00){\line(-4,3){7.00}}
\put(100.00,12.00){\line(5,2){13.00}}
\put(120.00,12.00){\line(1,0){20.00}}
\put(140.00,12.00){\line(-4,3){7.00}}
\put(120.00,12.00){\line(5,2){13.00}}
\put(112.89,17.22){\line(1,0){20.00}}
\put(132.89,17.22){\line(-4,3){7.00}}
\put(112.89,17.22){\line(5,2){13.00}}
\thinlines
\put(100.00,32.00){\line(1,0){20.00}}
\put(120.00,32.00){\line(-4,3){7.00}}
\put(100.00,32.00){\line(5,2){13.00}}
\put(120.00,32.00){\line(1,0){20.00}}
\put(140.00,32.00){\line(-4,3){7.00}}
\put(120.00,32.00){\line(5,2){13.00}}
\put(112.89,37.22){\line(1,0){20.00}}
\put(132.89,37.22){\line(-4,3){7.00}}
\put(112.89,37.22){\line(5,2){13.00}}
\put(100.00,7.00){\line(0,1){30.00}}
\put(113.00,13.00){\line(0,1){28.00}}
\put(126.00,19.00){\line(0,1){26.00}}
\put(120.00,7.00){\line(0,1){30.00}}
\put(140.00,7.00){\line(0,1){30.00}}
\put(133.00,14.00){\line(0,1){27.00}}
\put(141.00,12.00){\makebox(0,0)[lc]{$t=0$}}
\put(141.00,32.00){\makebox(0,0)[lc]{$t=\rm const$}}
\put(100.00,22.00){\makebox(0,0)[rc]{A}}
\put(140.00,22.00){\makebox(0,0)[lc]{A}}
\put(126.00,30.00){\makebox(0,0)[lc]{A}}
\put(120.00,22.00){\makebox(0,0)[lc]{C}}
\put(113.00,27.00){\makebox(0,0)[rc]{B}}
\put(133.00,27.00){\makebox(0,0)[lc]{D}}
\put(24.00,0.00){\makebox(0,0)[cc]{(a)}}
\put(65.00,0.00){\makebox(0,0)[cc]{(b)}}
\put(122.00,0.00){\makebox(0,0)[cc]{(c)}}
\end{picture}

\begin{caption}{(a) A tetrahedron is intrinsically flat everywhere except at the vertices. In (b) the tetrahedron is developed as a net of identical triangles in the plane by cutting it along the tree of geodesics AB, AC, AD. (c) The time development of the net consists of a analogous net of triangular prisms in Minkowski space. One prism can be obtained from an adjacent one by rotating about one of the common worldlines of the vertices.}
\end{caption}
\end{figure}

It is instructive to flatten out a regular polyhedron on a plane. This can be done (non-uniquely but for any $N$-particle universe) by cutting the polyhedron along a tree of geodesics connecting adjacent vertices and developing the figure in the plane so that pairs of faces are connected across at most one edge. The resulting ``net" of the polyhedron, together with gluing rules for the edges that were cut, is an equivalent way to describe the polyhedral surface. For example, we may cut the tetrahedron of Fig.~2a along $AB$, $AC$, and $AD$ to obtain Fig.~2b. The corresponding spacetime, unfolded, consists of the prisms over the faces of this figure (Fig.~2c). Adjacent faces in the 2-dimensional figure are related by a rotation (by $60^\circ$ in this case) about a common vertex, and adjacent prisms can be obtained by a similar rotation about the worldline of a common vertex. For example, prism ABC becomes prism BDC when rotated about worldline C in Fig.~2c.

A similar construction of the initial state of an $N$-particle universe works in the $\Lambda < 0$, negative curvature case. The only difference is that the angle sum of a face is less than that of the corresponding flat face, so that the angle deficit at the vertices is larger. The sum of the particle masses is therefore larger than $4\pi$, by an amount depending on the universe's size, similar to what one has in 3+1-dimensional time-symmetric Friedmann universes. For both $\Lambda = 0$ and $\Lambda < 0$, the universe with the smallest possible number of particles is not the surface of a finite solid: it has three particles and two (identical) triangular faces, which are glued together at their edges.\footnote{These may be regarded as ``Machian" results: the former is analogous to $GM/c^2R \sim 1$ for a closed universe of total mass $M$; and three particles are the minimum number needed to define an inertial frame in a 2+1-dimensional closed universe.}

\begin{figure}
\unitlength 1.00mm
\linethickness{0.4pt}
\begin{picture}(148.33,50.62)(-8,0)
\bezier{20}(11.67,30.00)(11.67,40.23)(20.84,46.04)
\bezier{20}(20.84,46.04)(30.00,50.62)(39.16,46.04)
\bezier{20}(39.16,46.04)(48.33,40.23)(48.33,30.00)
\bezier{20}(11.67,30.00)(11.67,19.77)(20.84,13.96)
\bezier{20}(20.84,13.96)(30.00,9.38)(39.16,13.96)
\bezier{20}(39.16,13.96)(48.33,19.77)(48.33,30.00)
\bezier{20}(61.67,30.00)(61.67,40.23)(70.84,46.04)
\bezier{20}(70.84,46.04)(80.00,50.62)(89.16,46.04)
\bezier{20}(89.16,46.04)(98.33,40.23)(98.33,30.00)
\bezier{20}(61.67,30.00)(61.67,19.77)(70.84,13.96)
\bezier{20}(70.84,13.96)(80.00,9.38)(89.16,13.96)
\bezier{20}(89.16,13.96)(98.33,19.77)(98.33,30.00)
\thicklines
\bezier{76}(19.44,26.22)(26.33,32.00)(27.56,42.22)
\bezier{20}(27.56,42.22)(30.00,41.33)(32.56,42.11)
\bezier{76}(32.56,42.11)(33.44,32.00)(40.67,26.33)
\bezier{20}(40.67,26.33)(39.11,24.89)(38.11,22.00)
\bezier{68}(38.11,22.00)(30.00,25.00)(21.89,22.11)
\bezier{20}(21.89,22.11)(21.00,24.44)(19.33,26.22)
\bezier{88}(77.90,43.14)(77.84,31.76)(68.24,26.30)
\bezier{64}(69.80,21.65)(77.84,23.73)(84.51,19.41)
\bezier{44}(88.43,21.96)(84.12,25.49)(88.43,29.02)
\bezier{60}(82.20,42.90)(82.75,31.76)(87.25,31.76)
\bezier{20}(67.84,26.08)(69.41,24.31)(70.0,21.57)
\bezier{12}(69.80,21.57)(68.82,22.94)(67.06,23.33)
\bezier{12}(67.06,23.33)(68.24,24.51)(67.84,26.27)
\bezier{16}(84.51,19.41)(84.71,22.16)(85.88,23.14)
\bezier{12}(85.88,23.14)(86.67,22.94)(88.43,21.96)
\bezier{24}(88.43,21.96)(85.69,21.96)(84.51,19.41)
\bezier{16}(77.84,43.14)(79.41,41.96)(80.98,41.57)
\bezier{8}(80.98,41.57)(81.57,41.96)(82.35,42.94)
\bezier{20}(82.35,42.94)(80.39,42.35)(77.84,43.14)
\bezier{12}(88.43,29.02)(87.06,30.00)(87.25,31.76)
\thinlines
\bezier{20}(67.25,23.33)(77.25,25.69)(85.69,29.61)
\bezier{80}(80.98,41.57)(80.98,30.00)(85.88,23.14)
\bezier{4}(87.25,31.76)(85.69,30.78)(85.69,29.61)
\bezier{4}(85.69,29.61)(86.27,28.63)(88.43,29.02)
\put(30.00,5.00){\makebox(0,0)[cc]{(a)}}
\put(80.00,5.00){\makebox(0,0)[cc]{(b)}}
\put(130.00,5.00){\makebox(0,0)[cc]{(c)}}
\bezier{20}(111.67,30.00)(111.67,19.77)(120.84,13.96)
\bezier{20}(120.84,13.96)(130.00,9.38)(139.16,13.96)
\bezier{20}(139.16,13.96)(148.33,19.77)(148.33,30.00)
\bezier{20}(111.67,30.00)(111.67,40.23)(120.84,46.04)
\bezier{20}(120.84,46.04)(130.00,50.62)(139.16,46.04)
\bezier{20}(139.16,46.04)(148.33,40.23)(148.33,30.00)
\bezier{76}(119.44,33.78)(126.33,28.00)(127.56,17.78)
\bezier{76}(132.56,17.89)(133.44,28.00)(140.67,33.67)
\bezier{68}(138.11,38.00)(130.00,35.00)(121.89,37.89)
\thicklines
\bezier{12}(128.60,45.20)(130.07,44.67)(131.53,45.20)
\bezier{36}(131.53,45.20)(133.13,41.47)(137.53,40.00)
\bezier{36}(137.53,40.00)(138.60,34.80)(141.93,32.40)
\bezier{44}(141.93,32.40)(139.93,27.47)(143.40,23.60)
\bezier{12}(143.40,23.60)(142.20,22.53)(142.07,20.93)
\bezier{40}(142.07,20.93)(137.53,21.73)(134.73,17.20)
\bezier{40}(134.73,17.20)(130.20,19.07)(125.67,17.20)
\bezier{36}(128.47,45.20)(126.87,41.47)(122.47,40.00)
\bezier{36}(122.47,40.00)(121.40,34.80)(118.07,32.40)
\bezier{44}(118.07,32.40)(120.07,27.47)(116.60,23.60)
\bezier{12}(116.60,23.60)(117.80,22.53)(117.93,20.93)
\bezier{40}(117.93,20.93)(122.47,21.73)(125.27,17.20)
\end{picture}

\begin{caption}{Constructing a black hole universe from right-angle hexagons with constant negative curvature. (a) The basic building block as it appears on the Poincar\'e disk. (b) In the Poincar\'e ball four building blocks fit smoothly together to make a black hole universe with tetragonal symmetry. The black hole horizons are at the throats of the four ``funnels" that protrude from the body of the universe. (c) The net of the universe on the Poincar\'e disk.}
\end{caption}
\end{figure}

\subsection{Universes with Black Holes}

For the initial state of an $N$-black-hole universe we need a figure analogous to the $N$-particle universe, but there must be gaps at the 
would-be vertices so the vertices are replaced by black hole horizons. The required figure to replace a triangular face is a right-angle, geodesic hexagon with alternate sides of 
equal length. This is shown in Fig.~3a in the Poincar\'e disk 
representation, where geodesics are arcs of circles that meet the disk's boundary at right angles. We call one set of three equal sides ``small", and the other 
set ``big." When four such hexagons are put together we obtain the 
surface of tetragonal symmetry of Fig.~3b. Pairs of big sides fit 
smoothly together to form the six edges, and three small sides at each 
would-be vertex, being perpendicular to the edges, join smoothly to make up the horizon of the black hole located there. (As in flat space, the ``creases" of the figure at the polygon edges are a feature only of the embedding.) The net of this solid is shown in Fig.~3c, which is analogous to Fig.~2b.

One can construct multi-black-hole universes with the symmetry of other 
regular polyhedra (or of the Archimedian 
solids, or of any $N$-particle universe). If the faces are triangular, they are replaced 
by hexagons as in the construction of the tetrahedral space in Fig.~3b; in general a face of a regular solid is replaced by an ``alternating-regular" polygon 
of twice the number of sides and right angles between adjacent sides. 
In a 2-dimensional constant negative curvature space such a figure can 
always be constructed, provided only that the length $b$ of the big sides 
is sufficiently 
large compared to the radius of curvature $\ell=\Lambda^{-1/2}$.
The minimum big side 
length occurs of course when the long and the short side lengths are equal, 
that is, for the completely regular right-angled $2N$-gon. 

On the other side of the horizons of a universe with $N$ horizons 
we may attach $N$ asymptotically AdS regions. Alternatively we 
may identify pairs of horizons to obtain $N/2$ wormholes that lead 
from one region of the universe to another along spacelike paths.  The genus of such a {\it compact} universe is $N/2$, hence the Gauss-Bonnet theorem 
implies for the area $A$ of our universe
\begin{equation}
A = \pi(N-2)/2\Lambda.
\end{equation}
The area depends only on $N$, not on the side lengths of the face 
polygons, and hence neither on the edge lengths nor on the horizon size 
$L$ of the black hole. Since the latter measures the mass parameter $m = \sqrt{L/\ell}$, $A$ is also independent of $m$; even extremal $m=0$ black holes can be used to build ``closed", finite-area universes.\footnote{The mass parameter $m$ of black holes measures a kind of bare mass, different from the particle mass. Particles are described by negative values of $m$, and the AdS vacuum has $m = -1$. Therefore Eq (1) contrasts with the condition on the total mass in a 2+1-dimensional dust-filled 
$k = +1$ Friedmann universe (to which the lattice universes are taken to be an 
approximation, in the spirit of Lindquist and Wheeler), where the 
constraint implies for a universe of radius $a$, area A and total mass $M = A\rho$ at the time-symmetric moment 
$$\dot a^2 = 0 = -1 - \left({A\over 4\pi\ell}\right)^2 + M.$$ 
so that $M$ must always exceed 1.}

\section{Klein Representation of Time-Symmetric Universes}

Geodesics and totally geodesic surfaces play an important role in the above treatment of universes put together out of flat or constant-curvature faces. The Klein disk and its generalization, in which geodesics of hyperbolic space ``look" straight, is a useful device to reduce many problems of hyperbolic space to those of flat space.

\begin{figure}
\unitlength 1.00mm
\linethickness{0.4pt}
\begin{picture}(144.36,60.47)(-12,10)
\bezier{120}(6.60,51.21)(6.60,54.23)(18.31,55.94)
\bezier{120}(18.31,55.94)(30.00,57.30)(41.69,55.94)
\bezier{120}(41.69,55.94)(53.40,54.23)(53.40,51.21)
\bezier{120}(6.60,51.21)(6.60,48.19)(18.31,46.47)
\bezier{120}(18.31,46.47)(30.00,45.12)(41.69,46.47)
\bezier{120}(41.69,46.47)(53.40,48.19)(53.40,51.21)
\bezier{264}(6.85,50.05)(30.00,19.95)(53.15,50.05)
\put(30.00,24.58){\vector(1,0){20.84}}
\put(30.00,46.58){\vector(0,1){12.73}}
\put(30.00,24.58){\vector(-1,-3){2.09}}
\thicklines
\put(48.52,28.05){\line(1,3){4.25}}
\put(6.85,28.05){\line(1,0){41.67}}
\thinlines
\put(6.85,28.05){\line(2,5){5.09}}
\put(11.83,40.79){\line(1,0){3.51}}
\put(52.77,40.79){\line(-1,0){7.72}}
\put(30.00,24.58){\line(0,1){3.47}}
\thicklines
\bezier{164}(11.48,47.93)(30.78,27.28)(35.01,46.00)
\thinlines
\put(31.93,32.30){\line(1,4){3.43}}
\put(22.67,33.84){\line(-4,5){11.26}}
\bezier{8}(22.67,33.84)(24.99,30.95)(27.30,28.05)
\bezier{16}(27.30,28.05)(28.65,26.32)(30.00,24.58)
\bezier{6}(28.07,32.88)(28.65,30.57)(29.22,28.05)
\bezier{12}(29.22,28.05)(29.62,26.32)(30.00,24.58)
\bezier{6}(31.93,32.30)(31.35,30.17)(30.96,28.05)
\bezier{12}(30.96,28.05)(30.38,26.32)(30.00,24.58)
\put(26.50,19.0){\makebox(0,0)[ct]{$x$}}
\put(51.99,24.58){\makebox(0,0)[lc]{$y$}}
\put(30.00,60.47){\makebox(0,0)[cb]{$t$}}
\bezier{12}(19.28,35.00)(19.28,33.32)(24.64,32.36)
\bezier{12}(24.64,32.36)(30.00,31.61)(35.36,32.36)
\bezier{12}(35.36,32.36)(40.72,33.32)(40.72,35.00)
\thicklines
\put(19.58,34.35){\line(6,-1){13.89}}
\thinlines
\put(28.07,32.94){\line(-1,4){1.07}}
\bezier{4}(40.65,35.08)(40.34,36.31)(38.80,36.77)
\bezier{4}(19.35,35.08)(19.66,36.31)(21.20,36.77)
\put(19.58,40.79){\makebox(0,0)[lb]{G}}
\put(16.11,31.53){\vector(2,1){4.63}}
\put(16.11,31.53){\makebox(0,0)[rt]{L}}
\bezier{20}(65.64,36.00)(65.64,27.98)(72.82,23.43)
\bezier{20}(72.82,23.43)(80.00,19.84)(87.18,23.43)
\bezier{20}(87.18,23.43)(94.36,27.98)(94.36,36.00)
\bezier{20}(65.64,36.00)(65.64,44.02)(72.82,48.57)
\bezier{20}(72.82,48.57)(80.00,52.16)(87.18,48.57)
\bezier{20}(87.18,48.57)(94.36,44.02)(94.36,36.00)
\put(65.00,44.00){\line(1,0){30.00}}
\put(65.00,44.00){\line(2,1){15.00}}
\put(95.00,44.00){\line(-2,1){15.00}}
\put(95.00,44.00){\line(-3,-5){15.00}}
\put(65.00,44.00){\line(3,-5){15.00}}
\put(65.00,44.00){\line(1,-6){2.83}}
\put(80.00,19.00){\line(-3,2){12.00}}
\put(95.00,44.00){\line(-1,-6){2.83}}
\put(80.00,19.00){\line(3,2){12.00}}
\put(74.00,23.00){\line(1,0){12.00}}
\put(93.00,23.00){\vector(-1,0){6.00}}
\put(93.00,23.00){\makebox(0,0)[lc]{H}}
\put(137.39,23.18){\line(-5,2){23.64}}
\put(137.39,23.18){\line(1,5){2.95}}
\put(130.00,52.73){\line(1,-4){7.39}}
\put(113.75,32.64){\line(4,5){16.07}}
\put(140.34,37.95){\line(-2,3){10.19}}
\bezier{20}(115.64,36.00)(115.64,27.98)(122.82,23.43)
\bezier{20}(122.82,23.43)(130.00,19.84)(137.18,23.43)
\bezier{20}(137.18,23.43)(144.36,27.98)(144.36,36.00)
\bezier{20}(115.64,36.00)(115.64,44.02)(122.82,48.57)
\bezier{20}(122.82,48.57)(130.00,52.16)(137.18,48.57)
\bezier{20}(137.18,48.57)(144.36,44.02)(144.36,36.00)
\bezier{50}(114.00,33.00)(127.00,36.33)(140.00,38.00)
\put(131.89,25.44){\line(5,6){3.61}}
\bezier{16}(135.78,29.67)(136.89,28.22)(138.00,26.78)
\bezier{24}(138.00,26.78)(134.89,26.11)(131.89,25.44)
\put(118.56,30.89){\line(-1,4){1.47}}
\put(138.56,40.56){\line(1,-5){1.11}}
\put(126.00,48.00){\line(3,-1){5.78}}
\put(131.78,46.07){\line(1,5){0.72}}
\put(132.50,49.67){\line(-4,-1){6.39}}
\bezier{12}(117.11,36.78)(119.33,34.44)(119.33,34.44)
\bezier{16}(119.33,34.44)(118.56,30.78)(118.56,30.78)
\bezier{16}(138.56,40.56)(137.11,39.00)(135.67,37.33)
\bezier{20}(135.67,37.33)(137.67,36.22)(139.67,35.11)
\put(130.33,31.22){\vector(4,-3){3.56}}
\put(130.22,31.22){\makebox(0,0)[rb]{H}}
\put(30.00,14.00){\makebox(0,0)[cc]{(a)}}
\put(80.00,14.00){\makebox(0,0)[cc]{(b)}}
\put(130.00,14.00){\makebox(0,0)[cc]{(c)}}
\end{picture}

\begin{caption}{(a) By central projection a geodesics G on the spacelike hyperboloid -- a space of constant negative curvature -- is mapped into a straight line segment L on the plane within the dotted Klein disk. (b) The Klein version of Figure 3(c) consists entirely of straight lines, which are here continued to their intersection outside the Klein disk. One of the horizons is indicated by H. (c) The four-black-hole universe, obtained by folding together the net of (b) in the Klein ball. The approximate location of the four horizons is indicated.}
\end{caption}
\end{figure}

The Klein model can be obtained in a similar way as the Poincar\'e model \cite{DB}, by using central rather than stereographic projection. The simplest example is 2-dimensional hyperbolic space, which can be embedded as the spacelike hyperboloid
$$t^2 - x^2 - y^2 = 1/\Lambda = R^2 $$
in Minkowski space. We project this surface by central projection on its horizontal tangent plane with rectangular coordinates $X$, $Y$ or polar coordinates $r$, $\phi$ where $X = r\sin\phi = x/t, \, Y = r\cos\phi = y/t$ (Fig.~4a). The metric induced on the hyperboloid by the Minkowski metric $ds_M^2 = -dt^2 + dx^2 + dy^2$ has constant negative curvature $1/R^2$ and takes the form in terms of the coordinates $r$, $\phi$ on the plane
\begin{equation}
ds^2 = {R^4\over (R^2 - r^2)^2}\, dr^2 + {R^2\over R^2-r^2}\, r^2d\phi^2 .
\end{equation}
The coordinate $r$ is restricted to $0 \leq r < R$, hence covers a disk of radius $R$, the Klein disk, whose boundary $r = R$ corresponds to infinity of hyperbolic space. Because those projection rays that intersect the disk in a line form a plane through the origin in Minkowski space, they intersect the hyperboloid in a geodesic. Therefore straight lines in the Klein disk represent geodesics of hyperbolic space. We note that, unlike the Poincar\'e map, the Klein map is not conformal, so that angles in the Klein disk are different (the smaller the closer to the boundary) than what they appear.

Consider now a regular hyperbolic polygon. It consists of equal geodesic edges. If we center it at the origin of the Klein disk, it looks exactly like the corresponding polygon in flat space, that is, like one having an equal number of vertices. Now expand this polygon, keeping it centered. Though the apparent vertex angles stay the same, the actual vertex angles decrease, in accordance with the metric (2), and reach zero when the vertices converge to the Klein disk's boundary. If this polygon is part of a polyhedron, the angle deficit at a vertex is now $2\pi$, so we can regard the universe as filled with particles of mass $2\pi$. Alternatively we can regard the limit as $m = 0$ black holes. If we increase the polygon further, so the vertices would be outside the Klein disk, we obtain the kind of face that can be put together into a black hole universe. In this sense black holes correspond to particles that lie ``beyond infinity" outside the Klein disk in the initial state. (As the metric (2) shows, radial distances outside the Klein disk are timelike, so these particles are not in any spacelike relation to the interior of the disk.)

We can, of course, unfold a whole hyperbolic polyhedron on the Klein disk (Fig.~4b), but because the central projection distorts shapes and distances as we move away from the center, the net will not bear much resemblance to the corresponding flat-space figure (cf. Figs.~4b and 2b). The resemblance is much better if one embeds the polyhedron itself in 3-dimensional hyperbolic space.

The Klein map can easily be extended to higher-dimensional spaces, 
for example by replacing the $d\phi^2$ of Eq (2) by the line element of a higher-dimensional sphere. Totally geodesic (hyper)surfaces map into (hyper)planes under the central projection. For example, in the Klein ball, a figure that looks like a Euclidean polyhedron, centered about the origin, represents a hyperbolic polyhedron with totally geodesic faces of constant intrinsic curvature, and its surface is the initial state of a $\Lambda < 0$ lattice universe with a point particle located at each vertex of the polyhedron. Again we can expand the polyhedron until its vertices fall outside of the Klein ball: the figure then represents the initial state of a universe containing black holes (Fig.~4c). The location of the horizons or throats is not easily identified in such a figure, but it is clear that there is an asymptotically anti-deSitter space on the ``other" side of each throat.

\section{Time Development of Time-symmetric Universes}

The initial data of the previous section can be continued in time, using normal coordinates. We consider the development into the future; corresponding statements hold for the past of the time-symmetric surface with ``collapse" replaced by ``big bang". For $\Lambda = 0$ the time-symmetric universe remains static, each face develops into a prism of Minkowski spacetime (Fig.~2c). For $\Lambda < 0$ the normal geodesics to any totally geodesic (that is, time-symmetric) spacelike surface in AdS space intersect at a conjugate point after the timelike interval $\pi R/2$, when the universe collapses to zero volume.\footnote{Unlike in the 3+1-dimensional case, the time to collapse is the same for all time-symmetric geodesic observers, whether near a black hole or not, and also for universes less regular than the lattice universes considered here. It is therefore more difficult in 2+1 dimensional universes to observe properties of black holes that depend on the observer outliving the black hole.} At an intermediate normal geodesic time $t$ the linear dimensions of the universe shrink uniformly by the factor $\cos(t/R)$, its intrinsic curvature increases by $\cos^{-2}(t/R)$, and its extrinsic curvature increases to $h = \tan(t/R)$, so as to maintain the constraint (or Gauss' {\it theorema egregium}),
$$^2R - h^2 = \Lambda .$$
Close to the collapse time these two curvatures dominate over $\Lambda$ (``curvature dominated" limit), and the behavior in the matter-free regions approaches that of a $\Lambda = 0$, flat-space model: at late times $t={1\over 2}\pi R - \epsilon$ the history of each face is a right pyramid in Minkowski space, with a base a surface of constant curvature a distance $\epsilon$ from the top vertex of the pyramid (Fig.~5a). The edges of the pyramid are the world lines of the point particles in the universe; they carry the angle deficit that occurs when the pyramids are assembled into the late history of the whole polyhedron. In the case of a black hole universe the base does not reach parts of the pyramid's sides that are near its edges. These edges (but not the faces) are spacelike (Fig.~5b) and carry a Lorentzian angle characteristic of the identification near the singularity of a BTZ black hole without angular momentum (see for example Figure 1b of Aminneborg et al \cite{B3}), or of Misner space \cite{M} near its non-Hausdorff singularity. Thus particles and black holes can be treated on a very similar footing, if we think of the black hole containing a spacelike, tachyonic particle.

\begin{figure}
\unitlength 0.80mm
\linethickness{0.4pt}
\begin{picture}(137.07,50.00)(-35,0)
\bezier{120}(9.93,20.00)(9.93,15.08)(23.47,12.29)
\bezier{120}(23.47,12.29)(37.00,10.08)(50.53,12.29)
\bezier{120}(50.53,12.29)(64.07,15.08)(64.07,20.00)
\bezier{300}(10.00,20.00)(37.00,46.00)(64.00,20.00)
\thicklines
\put(37.00,50.00){\line(-3,-5){15.00}}
\put(37.00,50.00){\line(1,-5){7.00}}
\put(37.00,50.00){\line(1,-2){9.50}}
\bezier{104}(22.00,25.00)(37.00,24.00)(44.00,15.00)
\bezier{96}(44.00,15.00)(41.00,33.00)(46.00,31.00)
\bezier{108}(46.00,31.00)(34.00,34.00)(22.00,25.00)
\thinlines
\bezier{120}(82.93,20.00)(82.93,15.08)(96.47,12.29)
\bezier{120}(96.47,12.29)(110.00,10.08)(123.53,12.29)
\bezier{120}(123.53,12.29)(137.07,15.08)(137.07,20.00)
\bezier{300}(83.00,20.00)(110.00,46.00)(137.00,20.00)
\thicklines
\put(110.00,50.00){\line(-5,-6){28.33}}
\put(110.00,50.00){\line(1,-4){10.10}}
\put(110.00,50.00){\line(5,-6){20.00}}
\bezier{172}(87.17,15.00)(106.61,29.44)(111.89,11.39)
\bezier{92}(123.00,12.50)(118.83,26.39)(126.89,27.78)
\thinlines
\put(81.89,15.83){\line(6,-1){5.28}}
\put(111.89,11.11){\line(5,-1){8.61}}
\put(73.00,50.00){\makebox(0,0)[cc]{collapse}}
\put(85.00,50.00){\vector(1,0){23.00}}
\put(61.00,50.00){\vector(-1,0){22.00}}
\put(37.00,5.00){\makebox(0,0)[cc]{(a)}}
\put(110.00,5.00){\makebox(0,0)[cc]{(b)}}
\bezier{25}(120.67,9.33)(121.83,10.83)(123.00,12.17)
\end{picture}

\begin{caption}{Collapse in Minkowski space, or in the velocity-dominated approximation in AdS space. The hyperboloid represents a constant time $\epsilon$ before the collapse. (a)~One face of a particle universe. The straight lines are the timelike worldlines of the particles. (b)~One face of a universe containing black holes. The straight lines are spacelike and correspond to the collapse of the individual black holes.}
\end{caption}
\end{figure}

The similarity between particles and black holes is brought into clearer focus in the Klein map of AdS space. This is obtained by a central projection of the 3-dimensional AdS surface on one of its tangent hyperplanes (Fig.~6a). The three-dimensional region covered by the projection is the interior of a timelike hyperboloid, the Lorentzian analog of the Klein ball (Fig.~6c). From Figure 6a one can see that, away from the center,  the projection reduces spacelike distances (as is the case for the Klein disk, which is contained as the central section $t=0$ of Fig.~6c), and magnifies timelike distances. Therefore the light cones become thinner near the boundary hyperboloid (Fig.~6b), and the interior covers only the interval $-\pi R/2 < t < +\pi R/2$ of AdS space. The AdS metric in terms of polar coordinates in the Klein hyperboloid can be obtained by an extension and analytic continuation of metric (2),
$$ ds^2 = {R^4\over (R^2-r^2)^2}\, dr^2 + {R^2 r^2\over R^2-r^2}\left(\cosh^2\psi d\phi^2 - d\psi^2\right), $$
where the coordinates $r, \psi, \phi$ are related to the Cartesian coordinates of the tangent hyperplane by $x = r\cosh\psi\cos\phi,\, y = r\cosh\psi\sin\phi,\, t = r\sinh\phi$. 

\begin{figure}
\unitlength 1.00mm
\linethickness{0.4pt}
\begin{picture}(163.40,57.67)(6,0)
\bezier{120}(52.00,47.33)(57.70,47.33)(60.93,38.16)
\bezier{120}(60.93,38.16)(63.48,29.00)(60.93,19.84)
\bezier{120}(60.93,19.84)(57.70,10.67)(52.00,10.67)
\put(62.00,31.00){\line(-3,2){17.00}}
\put(45.00,42.33){\line(-6,-1){36.00}}
\bezier{120}(6.28,23.00)(6.28,33.74)(11.64,39.84)
\bezier{120}(11.64,39.84)(17.00,44.65)(22.36,39.84)
\bezier{120}(22.36,39.84)(27.72,33.74)(27.72,23.00)
\bezier{120}(6.28,23.00)(6.28,12.26)(11.64,6.16)
\bezier{120}(11.64,6.16)(17.00,1.35)(22.36,6.16)
\bezier{120}(22.36,6.16)(27.72,12.26)(27.72,23.00)
\put(35.00,25.00){\circle*{0.80}}
\thicklines
\put(9.00,36.33){\line(4,-3){18.67}}
\put(27.67,22.33){\line(4,1){34.33}}
\put(51.33,28.00){\line(-4,3){16.67}}
\bezier{100}(39.00,37.17)(51.17,42.67)(46.83,31.33)
\thinlines
\bezier{150}(27.67,22.33)(34.50,33.67)(62.00,30.83)
\bezier{136}(51.83,10.67)(34.67,20.33)(24.83,9.33)
\bezier{184}(52.33,47.17)(42.83,27.17)(22.67,39.67)
\bezier{6}(46.83,31.33)(45.67,28.83)(43.83,26.33)
\bezier{72}(43.83,26.33)(40.00,20.50)(31.00,13.83)
\bezier{5}(31.00,13.83)(28.33,12.67)(25.83,12.17)
\bezier{112}(25.83,12.17)(14.50,11.33)(24.50,24.33)
\bezier{20}(24.67,24.67)(30.50,33.17)(39.00,37.00)
\bezier{40}(52.33,47.33)(48.83,47.83)(44.83,42.33)
\bezier{36}(44.83,42.33)(37.50,25.67)(46.83,13.00)
\bezier{24}(46.83,13.00)(49.00,10.50)(52.00,10.67)
\bezier{22}(35.00,25.00)(38.83,29.50)(42.83,34.50)
\put(42.83,34.33){\line(5,6){3.61}}
\bezier{68}(45.00,42.33)(41.83,37.50)(33.00,35.33)
\bezier{8}(33.00,35.33)(29.50,34.17)(25.80,34.80)
\bezier{68}(25.80,34.80)(18.33,35.00)(9.17,36.33)
\put(48.17,36.00){\makebox(0,0)[lc]{G}}
\bezier{40}(116.60,45.00)(116.60,48.18)(128.31,49.97)
\bezier{60}(128.31,49.97)(140.00,51.39)(151.69,49.97)
\bezier{40}(151.69,49.97)(163.40,48.18)(163.40,45.00)
\bezier{100}(116.60,45.00)(116.60,41.82)(128.31,40.03)
\bezier{120}(128.31,40.03)(140.00,38.61)(151.69,40.03)
\bezier{100}(151.69,40.03)(163.40,41.82)(163.40,45.00)
\bezier{40}(117.72,7.00)(117.72,3.15)(128.87,0.95)
\bezier{60}(128.87,0.95)(140.00,-0.77)(151.13,0.95)
\bezier{40}(151.13,0.95)(162.28,3.15)(162.28,7.00)
\bezier{20}(126.97,25.00)(126.97,26.95)(133.49,28.05)
\bezier{20}(133.49,28.05)(140.00,28.93)(146.51,28.05)
\bezier{20}(146.51,28.05)(153.03,26.95)(153.03,25.00)
\bezier{20}(126.97,25.00)(126.97,23.05)(133.49,21.95)
\bezier{20}(133.49,21.95)(140.00,21.07)(146.51,21.95)
\bezier{20}(146.51,21.95)(153.03,23.05)(153.03,25.00)
\thicklines
\bezier{52}(85.24,33.48)(90.00,34.66)(94.76,33.48)
\bezier{52}(94.76,33.48)(98.57,32.00)(94.76,30.52)
\bezier{52}(94.76,30.52)(90.00,29.34)(85.24,30.52)
\bezier{52}(85.24,30.52)(81.43,32.00)(85.24,33.48)
\thinlines
\put(56.67,24.67){\vector(-3,2){4.33}}
\put(57.00,24.33){\makebox(0,0)[lc]{L}}
\put(70.00,5.00){\line(1,1){40.00}}
\put(110.00,5.00){\line(-1,1){40.00}}
\put(110.00,8.83){\line(-4,5){28.93}}
\put(110.00,7.67){\line(-3,5){22.40}}
\put(109.00,5.00){\line(-2,5){16.00}}
\put(109.00,5.00){\line(-2,5){16.00}}
\put(106.63,5.00){\line(-1,4){10.00}}
\put(102.00,5.00){\line(0,1){40.00}}
\put(70.00,8.83){\line(4,5){28.93}}
\put(70.00,7.67){\line(3,5){22.40}}
\put(71.00,5.00){\line(2,5){16.00}}
\put(73.38,5.00){\line(1,4){10.00}}
\put(78.00,5.00){\line(0,1){40.00}}
\put(110.00,41.17){\line(-4,-5){28.93}}
\put(110.00,42.33){\line(-3,-5){22.40}}
\put(109.00,45.00){\line(-2,-5){16.00}}
\put(109.00,45.00){\line(-2,-5){16.00}}
\put(106.63,45.00){\line(-1,-4){10.00}}
\put(70.00,41.17){\line(4,-5){28.93}}
\put(70.00,42.33){\line(3,-5){22.40}}
\put(71.00,45.00){\line(2,-5){16.00}}
\put(73.38,45.00){\line(1,-4){10.00}}
\bezier{104}(118.00,43.00)(136.13,25.20)(118.00,7.00)
\bezier{104}(162.00,43.00)(143.87,25.20)(162.00,7.00)
\thicklines
\put(90.00,25.00){\line(1,1){6.00}}
\put(90.00,25.00){\line(-1,1){6.00}}
\put(125.49,24.09){\line(6,-1){17.23}}
\put(142.72,21.22){\line(5,2){11.79}}
\thinlines
\put(154.51,25.94){\line(-6,1){17.23}}
\put(137.28,28.81){\line(-5,-2){11.79}}
\put(125.49,44.09){\line(6,-1){17.23}}
\put(142.72,41.22){\line(5,2){11.79}}
\put(154.51,45.94){\line(-6,1){17.23}}
\put(137.28,48.81){\line(-5,-2){11.79}}
\put(125.47,48.00){\line(0,-1){41.87}}
\put(142.53,44.93){\line(0,-1){41.87}}
\put(137.47,52.67){\line(0,-1){41.87}}
\put(154.53,49.60){\line(0,-1){41.87}}
\put(137.47,28.80){\line(2,-3){5.07}}
\put(154.53,32.80){\line(-1,-2){4.27}}
\put(150.27,24.27){\line(-1,1){2.80}}
\put(147.47,27.07){\line(5,4){7.07}}
\bezier{16}(147.60,27.07)(148.93,25.73)(150.27,24.27)
\put(148.00,27.60){\line(0,-1){1.07}}
\put(149.07,28.27){\line(0,-1){2.67}}
\put(150.00,29.07){\line(0,-1){4.40}}
\put(150.93,29.87){\line(0,-1){4.13}}
\put(152.00,30.67){\line(0,-1){2.80}}
\put(153.07,31.47){\line(0,-1){1.47}}
\put(125.00,38.00){\makebox(0,0)[cc]{A}}
\put(143.00,35.00){\makebox(0,0)[cc]{B}}
\put(137.00,41.00){\makebox(0,0)[cc]{C}}
\put(155.00,39.00){\makebox(0,0)[cc]{A}}
\put(137.47,48.80){\line(2,-3){5.07}}
\put(35.00,-5.00){\makebox(0,0)[cc]{(a)}}
\put(90.00,-5.00){\makebox(0,0)[cc]{(b)}}
\put(140.00,-5.00){\makebox(0,0)[cc]{(c)}}
\thicklines
\bezier{20}(12.00,34.50)(37.00,35.00)(26.00,24.00)
\end{picture}
\vskip5\unitlength

\begin{caption}{(a) 1+1 AdS space, embedded in 1+2 Minkowski space, is projected from the central black dot of the figure onto a plane (Klein map). The projection covers a region bounded by two hyperboloids, only one of which is shown (dotted). Any geodesic G of AdS space (the intersection of a plane through the center with the hyperboloid) is mapped on a straight line L. (b) The null geodesics of AdS space in the Klein map show the behavior of the light cones, such as the heavily drawn central one. (c) The net of a simple black hole universe in the 2+1-dimensional Klein map.}
\end{caption}
\end{figure}

A BTZ black hole may be represented in the Klein map as a slab of the timelike hyperboloid bounded by two timelike planes, which are to be identified. The two asymptotically AdS infinities are two strips cut out of the boundary hyperboloid by the planes. The two planes intersect at positive and negative timelike infinity, where the singularity is located. Similarly, in the Klein map the time-continuation of the faces of a polyhedral time-symmetric particle universe with $\Lambda < 0$ become prisms, as in Fig.~2c, representing the unfolded net of a polygonal universe. For the case of a black hole universe the net looks very similar, as in Figure 6c, where for simplicity only the ``trihedral" 3-particle universe is shown rather than the tetrahedral one of Fig.~2c. It also consists of prisms, but the vertical edges initially lie {\em outside} the boundary hyperboloid. Due to the tilting of the lightcones and their aligning with the boundary, these edges enter the interior as {\em spacelike} geodesics -- the tachyonic singularities of the black holes. These singularities do not lie entirely in the map's future infinity, because the black holes are moving (collapsing) in this frame. By a suitable AdS isometry one could move any one of the singularities to the Klein map's future infinity, to look like the singularity of the BTZ black hole.
Alternatively, by a timelike isometry one could move the entire singularity to a finite region of the Klein map.
For a particle universe the singularity is the single point where all the particle worldlines meet, and for a black hole universe it is a set of spacelike geodesics that meet at a point. The striped region in Fig.~6c is part of the true horizon of black hole A, and the region to the right of it is the asymptotically AdS ``exterior" of that black hole.

The state of an AdS lattice universe at a time slice $t \neq 0$ provides initial values with non-vanishing extrinsic curvature, in which the particles or black holes are uniformly moving together or apart. This then provides states with a momentum variable conjugate to the one ``position" variable, such as edge length or total volume, necessary to specify a time-symmetric state. 

By contrast, for Minkowski-space time-symmetric lattice universes all the slices $t = $const. are alike. To obtain initial states with non-vanishing extrinsic curvature we should therefore construct a net of pyramids like those in Figure 5, which is valid for all $\epsilon$ in this case. If the edges in this Minkowski-space construction are spacelike, we obtain one or several black holes whose intrinsic geometry is like that of BTZ black holes for a suitable $\Lambda < 0$ -- but the extrinsic curvature is too large for true BTZ black holes. In fact, one knows that there are no true black holes in locally flat spacetime, and we have them here only because of our somewhat looser concept of black holes in closed universes.

\section{Universes with Angular Momentum}

A regular polygon can remain regular not only when it expands or contracts, it can also rotate. We will consider the case of rotating particles in Minkowski space, and rely on the Klein map to translate these results to the case of black holes and AdS space.

Our rotating particles have to move on geodesics in Minkowski space, and in order to have nonzero angular momentum (about the center of a polyhedral face) they must be mutually skew, avoiding common intersection at a point. For a pair of skew world lines there is a distinguished epoch, namely the ``moment of closest approach", represented by the spacelike geodesic that is normal to both of them. (In Euclidean space this would also be the shortest distance between them.) We will call these geodesics ``connectors." A connector corresponds to a side of a polygonal face, and it is dual to the world lines it connects: the normal geodesic to a pair of ``adjacent" connectors is the world line of the common vertex. Because the connectors are skew, like the world lines, they do not in fact form a closed polygon. There is no smooth spacelike surface in such a universe, a circumstance familiar from the G\"odel universe \cite{G}. For each polygonal face we could of course find spacelike surfaces, for example those corresponding to the center of momentum frame, but these do not assemble into a continuous polyhedron. (The worldlines of the total momentum of each face correspond to the conjugate polyhedron, where faces are replaced by vertices and vice versa, but those worldlines are also mutually skew.) Therefore we assemble the universe out of whole spacetime regions, corresponding to the history of a face.

The assembly instructions for our spacetime are a sequence of isometries that map wordlines into worldlines according to the same scheme as for the net of the polyhedron, for example that in Fig.~2c. The boundaries of the regions do not matter, as long as the regions are large enough so that the isometries will generate overlaps. One can also define boundaries so that the regions fit together exactly and smoothly, like the plane boundaries of the prisms and pyramids of Figures 2, 5, and 6. These boundaries, containing the skew world lines, cannot of course be plane, but for faces with an even number of worldlines the boundary between one pair of adjacent worldlines can be rather arbitrary, for the other boundaries can then be consistently generated by the isometries. For an odd number of worldlines the only choice appears to be ruled surfaces between adjacent worldlines that include the connector.

\begin{figure}
\unitlength 1.00mm
\linethickness{0.4pt}
\begin{picture}(129.66,73.17)(-20,0)
\put(66.01,60.91){\line(-3,-4){37.84}}
\put(27.63,21.83){\line(-1,5){9.24}}
\put(31.18,42.67){\line(0,0){0.00}}
\bezier{8}(27.63,21.83)(27.75,21.35)(28.34,21.83)
\bezier{8}(26.92,11.05)(26.92,10.46)(27.86,10.34)
\put(22.42,48.60){\line(5,-4){20.49}}
\put(23.86,48.76){\line(4,-3){20.02}}
\put(23.08,48.98){\oval(1.25,0.71)[t]}
\bezier{12}(42.79,32.25)(43.62,32.72)(43.62,33.55)
\put(24.91,41.61){\line(6,1){4.85}}
\put(51.79,46.82){\line(-5,-1){20.02}}
\put(24.91,42.56){\line(6,1){4.03}}
\put(30.47,43.62){\line(5,1){21.20}}
\put(26.80,10.93){\line(3,4){15.99}}
\put(43.74,33.55){\line(3,4){21.04}}
\bezier{8}(65.89,61.04)(65.06,61.15)(64.82,61.75)
\bezier{8}(64.82,61.75)(65.65,61.63)(66.01,61.04)
\put(28.34,21.78){\line(-1,6){4.21}}
\put(23.86,48.75){\line(-1,6){3.29}}
\put(19.46,68.10){\circle{2.37}}
\put(58.47,16.21){\circle{2.38}}
\put(47.57,72.61){\line(1,-6){4.58}}
\put(48.41,72.75){\line(1,-5){5.15}}
\put(57.04,16.17){\line(-1,6){3.53}}
\put(59.55,16.45){\line(-1,5){5.63}}
\bezier{8}(47.57,72.46)(47.71,73.17)(48.41,72.61)
\put(49.90,39.59){\line(6,-1){4.42}}
\put(54.16,37.54){\line(-1,0){5.37}}
\put(48.48,37.54){\line(1,0){5.37}}
\put(53.06,39.12){\line(-1,6){0.61}}
\put(54.40,38.25){\oval(1.11,1.42)[r]}
\put(97.63,21.83){\line(-1,5){9.24}}
\bezier{8}(97.63,21.83)(97.75,21.35)(98.34,21.83)
\put(89.50,68.10){\circle{2.37}}
\put(128.47,16.21){\circle{2.38}}
\bezier{8}(117.00,72.46)(117.71,73.17)(118.41,72.61)
\put(97.67,10.00){\line(1,2){29.67}}
\put(98.50,21.83){\line(-1,6){7.72}}
\put(96.33,35.83){\line(6,5){20.83}}
\put(96.33,35.00){\line(5,4){19.33}}
\put(103.67,22.16){\line(3,4){18.75}}
\put(122.33,48.00){\line(-4,-5){14.80}}
\put(94.00,49.50){\line(2,-1){11.83}}
\put(106.83,43.16){\line(2,-1){4.33}}
\put(116.00,38.50){\line(2,-1){8.50}}
\put(94.17,48.16){\line(2,-1){10.33}}
\put(105.67,42.33){\line(2,-1){4.83}}
\put(115.17,37.66){\line(2,-1){9.33}}
\put(127.17,16.00){\line(-1,6){2.81}}
\put(124.00,34.66){\line(-1,6){1.97}}
\put(121.83,47.50){\line(-1,6){1.33}}
\put(120.00,58.83){\line(-1,5){2.70}}
\put(129.50,16.83){\line(-1,5){8.07}}
\put(120.83,60.50){\line(-1,5){2.40}}
\bezier{4}(122.17,47.83)(122.83,47.50)(122.33,47.00)
\bezier{8}(124.50,34.16)(125.50,33.50)(124.67,32.83)
\put(125.83,70.00){\line(-1,-2){29.50}}
\bezier{12}(96.17,11.00)(96.17,9.83)(97.50,10.00)
\bezier{8}(125.83,70.16)(126.50,69.33)(127.33,69.66)
\bezier{8}(127.33,69.66)(126.83,70.33)(125.83,70.16)
\bezier{8}(112.24,39.00)(113.24,38.56)(114.24,38.00)
\bezier{12}(112.79,40.11)(113.91,39.56)(115.13,39.00)
\put(42.79,5.00){\makebox(0,0)[cc]{(a)}}
\put(109.13,5.00){\makebox(0,0)[cc]{(b)}}
\end{picture}

\begin{caption}{(a) Three lines in Euclidean space that are mutually skew are drawn as three approximately vertical rods. The three approximately horizontal rods represent the unique lines that are orthogonal to both of the skew lines connected by them. The Euclidean sense of orthogonality is used to make the figure easier to visualize. (b) The analogous (but more confusing) figure in Minkowski space shows three timelike, non-intersecting worldlines, pairwise connected by "connectors" that are Minkowski-orthogonal to the worldlines at both ends.}
\end{caption}
\end{figure}

What should be the isometries that generate a new region of the net from a given one, to replace the rotation about a vertex world line of the time-symmetric case? It should leave that vertex wordline fixed and map one adjacent worldline into the other adjacent one. Because of the duality between worldlines and connectors this is equivalent to demanding that one connector at the vertex worldline should map into the other one. It should be clear from Figure 7 that this involves a rotation about the world line {\em and} a translation along the worldline.

Now consider what happens when we derive the properties of the particle by finding the total isometry at that vertex as we go once around the vertex through all the faces that meet there: In the spacetime assembled from the net we come back to the same point, but in Minkowski space we will have rotated around the vertex worldline and also translated in the direction of the worldline, like ascending an ``Escher staircase" that comes back to the original step after $n$ steps up. This 
translation signals that the worldline is a source of angular momentum as well as mass \cite{DJT}. Thus, if there is orbital angular momentum in a face, the particles must necessarily have spin angular momentum; and vice versa it is possible to put spinning particles into the closed universe only if the particles also have angular momentum about each other.\footnote{This can of course be regarded as another ``Machian" result for these closed universes.}

The construction of the analogous AdS and black hole universes is now straightforward, thanks to the Klein map that allows us to translate from a Minkowski space diagram to an AdS diagram. In broad outline the interpretation is the same as in Minkowski space, but the details, such as the values of masses and angular momenta, are different. Also, skew geodesics within the Klein hyperboloid do not intersect at Klein infinity but remain skew, so the rotating universe does not collapse to zero volume. In 2+1 dimensions angular momentum can stop collapse, so that the universe can last forever.

In the case of black holes the particle wordlines are replaced by spacelike skew worldlines, with an associated Lorentz angle measuring the black hole mass, and a spacelike translation measuring the black hole spin \cite{DBA}. BTZ black holes  with intrinsic angular momentum contain a region of closed timelike lines. This region is usually considered an unphysical, singular part of spacetime.\footnote{More accurately, a sufficiently large part of the region of closed timelike curves is eliminated from spacetime so that the remainder does not support any closed timelike lines. If one includes in the unphysical closed timelike lines not only those associated with one black hole, but also ones that travel all around the universe, there may not be any physical spacetime left.} If we take the same view for black hole universes then angular momentum of black holes does not stop the collapse, and the physical universe containing spinning black holes has only a finite lifetime. 

\section{Conclusion}
We have discussed some particular spacetimes in 2+1-dimensional gravity. In a sense these are the simplest after the point particle and the BTZ black hole solutions, due to their symmetry. This symmetry makes it natural to describe the spacetime in terms of a possibly large number of neighborhoods (faces of a polyhedron), using a combinatorial approach. Because geodesics and totally geodesic surfaces play an important role in this approach, we found the generalization of the Klein map to AdS space useful for transferring results from locally flat (Minkowski) to locally AdS spacetime. Identified flat spacetimes, typically containing point particles, have analogs in identified AdS spacetimes, with identical topological but different metrical properties. Static flat spacetimes correspond to time-symmetric, expanding and recontracting, models under this map.

The Klein map also provides Minkowski space analogs of AdS black holes. In this description a BTZ black hole is characterized by a straight line that is not contained entirely within the ``Klein hyperboloid," and enters the latter at the endpoint of future null infinity. Multi-black-hole spacetimes can be treated in this way, and spin angular momentum of black holes corresponds under the Klein map to  the particle spin of the straight line associated with the black hole.

\section*{Acknowledgments}
I am grateful to the Erwin Schr\"odinger Institute of Vienna, where this work was begun, and to the American Institute of Mathematics, where it was finished, and where I was afforded the opportunity for useful discussion with Vince Moncrief.

\end{document}